\documentclass{elsart}
\usepackage{epsfig}
\usepackage{amsmath}

\usepackage{amssymb}

\begin{document}

\begin{frontmatter}



\title{Self-similar solutions and collective coordinate methods
 for Nonlinear Schr\"{o}dinger Equations}


\author{V\'{\i}ctor M. P\'erez-Garc\'{\i}a},

\address{Departamento de Matem\'aticas, Escuela T\'ecnica
Superior de Ingenieros Industriales,
  Universidad de Castilla-La Mancha, 13071 Ciudad Real, Spain}

\begin{abstract}
In this paper we study the phase of self-similar solutions to
general Nonlinear Schr\"odinger equations. From this analysis we
gain insight on the dynamics of nontrivial solutions and a deeper
understanding of the way collective coordinate methods work. We
also find general evolution equations for the most relevant
dynamical parameter $w(t)$ corresponding to the width of the
solution. These equations are exact for self-similar solutions
 and provide a shortcut to find approximate evolution
equations for the width of non-self-similar solutions similar to
those of collective coordinate methods.
\end{abstract}

\end{frontmatter}

\section{Introduction}
\label{I}

The nonlinear Schr\"{o}dinger equation (NLSE) in its many versions
is one of the most important models of mathematical physics, with
applications to different fields such as plasma physics, nonlinear
optics, water waves and biomolecular dynamics, to cite only a few
cases. In many of those examples the equation appears as an
asymptotic limit for the  slowly varying envelope of a dispersive
wave propagating in a nonlinear medium \cite{VVz,Sulem}. A new
burst of interest on nonlinear Schr\"{o}dinger equations has been
triggered by the experimental achievement of Bose-Einstein
condensation (BEC) using ultracold neutral bosonic gases
\cite{Siempre,Siempre2}.

Many different forms of the Nonlinear Schr\"odinger equation
appear in practical applications but many of them are included in
the following family of equations
\begin{equation}\label{NLS}
i \frac{\partial u}{\partial t} = -\frac{1}{2}\triangle u + V(x,t)
u + g(|u|^2; t) u,
\end{equation}
where $u(x,t)$ is a complex function to be determined, $x\in
\Omega \subset \mathbb{R}^n$, $V(x,t)$ is a real function called
the \emph{potential} and $g$ is a real function satisfying $g(y;
t) \rightarrow 0, y\rightarrow 0$ accounting for the
nonlinearities. In many physical applications $\Omega \equiv
\mathbb{R}^n$ and $u$ has at least finite $L^2$ and $H^1$ norms in
the spatial variables because of its physical meaning of
\emph{mass} and \emph{energy} of the solutions. Most of the
results to be presented in this paper can be extended without
problems to the so-called vector case, when $u = (u_1,...,u_M)$,
however, to keep the formalism as simple as possible we
concentrate here on the scalar case.

 Many questions may be posed from the mathematical point of view
 on the properties of solutions of Eqs. (\ref{NLS}). Results are available for
 local and global existence, asymptotic behavior, etc ...
 \cite{Sulem,asy1,asy2}. In some very restricted one dimensional situations
 it is even possible to find the analytical expression of the
 solutions by the use of the so-called inverse scattering transform or
 equivalent methods \cite{inverse}.
  Finally, the so-called moment method allows in some specific cases to obtain rigorous
 results for the evolution of integral quantities related to the
 solution of Eq. \eqref{NLS} \cite{moment1,moment2,moment3}.

 However, very little is known (rigorously) on the dynamics of solutions which are
 asymptotically non-stationary (e.g. when either the conditions discussed
 in Refs. \cite{asy1,asy2} do not hold or there are more complex
 non-autonomous situations, etc).

For those situations it is customary in the applied sciences to
use the so called time-dependent variational method, collective
coordinates method or averaged-Lagrangian method
\cite{Angel,Borisreview} to obtain some insight on the behavior of
the solutions. All of the names refer to the same idea, which
consists on rewriting Eq. (\ref{NLS}) as a variational problem on
the basis of the Lagrangian density
so that the problem of solving Eq. (\ref{NLS}) is transformed into
the one of finding $u(x,t)$ such that the action
\begin{equation}\label{density}
S = \int \mathcal{L}(u,u^*,\nabla u, \nabla u^*,x,t) d^nx \, dt
\end{equation}
be one extremum. Of course, this problem is in principle as
complicated as that of solving the original nonlinear
Schr{\"o}dinger equation \eqref{NLS}. The idea of the method of
collective coordinates is to restrict the analysis to a specific
family of functions, that is to find the extremum of
\eqref{density} over a specific family of functions $u(x,t) =
\varphi(x,q_1(t),...,q_K(t);t)$ with fixed $\varphi$ and letting
$q_1(t),...,q_K(t)$ be unknown functions containing the dynamics
of $\varphi$. Once the test function is chosen one may explicitly
compute an averaged Lagrangian
\begin{equation}
L(t) = \int \mathcal{L}(\phi,\phi^*,\nabla \phi, \nabla
\phi^*,x,t) d^nx, \end{equation} and the problem becomes a
finite-dimensional one for the unknowns $q_1,...,q_K$. Thus, the
evolution equations for the parameters are obtained by solving the
Euler-Lagrange equations
\begin{equation}
\frac{d}{dt} \left( \frac{\partial L}{\partial
 \dot{q_j}}\right)
  - \frac{\partial L}{\partial q_j} = 0
 \; , j=1,...,K.
\end{equation}
The accuracy of the results depends crucially on the proper choice
of the family of test functions $\varphi$. In practice, this
choice requires some knowledge of the solution obtained either
from numerical simulations, asymptotic behaviors, or on the basis
of more or less heuristic reasoning. Also the type of solutions
studied usually have a simple form to keep the complexity of the
calculations under control.

 There is an immense body of references on
the application of the collective coordinate method to many
specific applications, normally from a heuristic perspective. Two
panoramas for its application in different fields can be seen in
\cite{Angel} and \cite{Borisreview}.

One drawback of the method is that in the framework of Nonlinear
Schr\"odinger systems it is usually possible to guess an
approximate form for the amplitude of the solution, but the choice
of the phase is more difficult to justify. Usually a simple form
is chosen based on the experience of the linear Schr\"odinger
equations as will be discussed later.

In this paper we propose a way to choose the phase systematically
by solving exactly the phase equation corresponding to
self-similar solutions. We also discuss how to use this phase to
construct simple evolution equations for the parameters which
improve the variational equations. In this analysis it will be
also clear why the usual form of the method of collective
coordinates is able to provide reasonable results for the
dynamics.

\section{Formulation of the problem}

 We start our analysis from Eq. (\ref{NLS}) and
 split $u$ into amplitude and phase by writing $u(x,t) =
a(x,t) e^{i\phi(x,t)}$. After substitution in Eq. \eqref{NLS} we
obtain the following coupled system of equations
\begin{subequations}
\label{eqaf}
\begin{eqnarray}
\frac{\partial a}{\partial t} & = & - \left(\nabla
a\right)\left(\nabla \phi\right)-\frac{1}{2}a
\triangle \phi, \label{eq1:1} \\
\rho \frac{\partial \phi}{\partial t} & = & \frac{1}{2} \triangle
a - \frac{1}{2} (\nabla \phi)^2 a - V(x,t) a - g(a,t)\rho.
\label{eq:2}
\end{eqnarray}
\end{subequations}

From now on we concentrate on non-propagating solutions of Eq.
(\ref{NLS}), i.e., those for which the ``centrum" of the
wavepacket defined as $X_j(t) = \int x_j (a(x,t))^2 d^nx$
satisfies $X_j(t) = 0$. Because of the Ehrenfest theorem
\cite{Mate,VVK}, for any type of nonlinear term $g$ and for
cuadratic polynomial potentials, the evolution of this quantity
can be decoupled exactly from the ``internal" dynamics of the
solution so that starting from solutions $u(x,t)$ with $X_j(t)
\neq 0$ we may construct new solutions $u(x-X_j(t),t)$ satisfying
$X_j(t) = 0$ as discussed in \cite{VVK}. Thus for the above
mentioned type of problems (which include many problems of
practical interest) we may concentrate on the analysis of this
type of solutions without loss of generality. For nonlinear
Schr\"odinger equations with algebraic potentials or polinomial
potentials of degree higher than two our analysis will be
applicable to non-propagating solutions.

\subsection{The phase of self-similar radially symmetric
solutions}

As discussed above, the key idea of the method of collective
coordinates is to restrict the analysis to a particular set of
functions $\varphi(x,q_1(t),...,q_M(t); t)$. In this paper we will
call quasi-solutions to these functions which are not true
solutions of the original equations \eqref{NLS} but in some sense
provide some estimates for the dynamics provided the parameters
$q_j(t)$ are chosen optimally and the choice of the set of trial
functions is appropriate. In the case of Schr\"odinger equations,
the time-dependent variational method proceeds usually by choosing
quasi-solutions of the form
\begin{equation} \label{ccc}
\varphi = A(t) \rho(r/w(t)) e^{i\beta(t) r^2}.
\end{equation}
Thus, when non-propagating symmetric systems are considered there
appear usually three free parameters $A(t), w(t), \beta(t)$ to be
determined (they have the physical meaning of amplitude, width and
chirp of the solution) once the motion of the center has been
decoupled. The consideration of non-symmetric systems over $n$
spatial dimensions leads to $2n+1$ independent parameters and can
be handled in a similar way. The specific choice for the phase
$\beta(t) r^2$ is believed to be a good one on the basis of what
it happens in linear systems and previous experience with the
equation. In fact, this term is chosen to be quadratic by analogy
with the Optics of gaussian beams, ruled by linear Schr\"odinger
equations. In that field it is well known that gaussian beams must
include a cuadratic phase term to be exact solutions of the linear
propagation equations. One of the results of this paper will be to
provide an understanding of why this approximation works so well
for the nonlinear case.

The main point of our analysis is to study self-similar solutions
of the form
\begin{equation} \label{selfsimm}
u(x,t) = A(t) \rho(r/w(t)) e^{i\phi(r,t)}.
 \end{equation}
 being $A(t)$, $w(t)$, $\rho$ and $\phi(r,t)$ real functions.

 When self-similar solutions to Eq.
(\ref{NLS}) exist our Eq. (\ref{selfsimm}) will provide exact
solutions. When self-similar solutions do not exist we will
understand Eq. (\ref{selfsimm}) as an ansatz approximating the
true solutions in a sense similar to that of the quasi-solutions
of the method of collective coordinates. The point we will develop
in what follows is that the phase can be found in closed form
independently of the specific form of the amplitude $\rho$.

One key feature of NLS equations given by Eq. (\ref{NLS}) is the
conservation of the $L^2$-norm $\| u \|_2 = \left(\int |u(x,t)|^2
d^nx\right)^{1/2}$ under time evolution. For self-similar
solutions and choosing the norm value to be equal to one (this can
be done without loss of generality by scaling appropriately the
nonlinear term) this conservation law becomes
\begin{equation}
A^2(t) \int_{\mathbb{R}^n} (\rho(r/w))^2 d^nx = 1.
\end{equation}
Let us choose $\rho$ such that $\|\rho\|_2=1$ (this can be done
without loss of generality and determines uniquely $A(t)$), then
\begin{equation} A^2 w^n = 1,\label{normac}\end{equation} thus
\begin{equation}\label{dotA}
\dot{A} = -\frac{n}{2}\dot{w}.
\end{equation}

The usual situation in applied sciences corresponds to functions
for which not only the $L^2$-norm is well defined (and conserved),
but also the $H^1$ norm is well defined. In fact, the gradient
term $\int |\nabla u|^2 d^nx$ is sometimes called the kinetic
energy, which must be finite for physically meaningful solutions.
Thus, we will look for self-similar solutions of the form
\eqref{selfsimm} with finite $H^1$ and $L^2$ norms. Situations
where blow-up occurs require an specific analysis near the blow-up
point where $\int |\nabla u|^2 \rightarrow \infty$ \cite{Sulem}.

Substituting (\ref{selfsimm}) into (\ref{eq1:1}) gives
\begin{multline}
\dot{A} \psi(r/w(t))  - \frac{\dot{w}}{w^3} A(t) \psi_r(r/w(t)) =
\\ - A(t) \rho_r(r/w) \phi_r(r/w)-\frac{1}{2}A(t) \rho(r/w)
\left[\phi_{rr} + \frac{n-1}{r}\phi_r\right],\label{multi}
\end{multline}

 We now define a new variable $q = r/w(t)$ and then Eq.
(\ref{multi}) becomes, after using \eqref{dotA}
\begin{equation}
\phi_{qq} + \left(\frac{n-1}{q} + \frac{2\rho_q}{\rho}\right)
\phi_q =
2\dot{w}w\left(\frac{n}{2}+\frac{\rho_q}{\rho}q\right).\label{phaseq0}
\end{equation}
It is remarkable that this is a linear second order equation whose
general solution can be easily found. Defining $v(q) =
\phi/(2w\dot{w})$ we get
\begin{equation}
v_{q} + \left(\frac{n-1}{q} + \frac{2\rho_q}{\rho}\right) v =
\frac{n}{2}+\frac{\rho_q}{\rho}q.\label{phaseq}
\end{equation}
The formal general solution of Eq. (\ref{phaseq}) is
\begin{equation}
v(q) = \frac{C_2(t)}{q^{n-1}\rho(q)^2} + \frac{q}{2}.
\end{equation}
for any arbitrary function $C_2(t)$. Thus, integrating we get
\begin{equation}\label{phasefin}
\phi(q,t) = 2w\dot{w} \left[C_1(t) + C_2(t) \int^q
\frac{ds}{s^{n-1}\rho(s)^2} + \frac{q^2}{4}  \right].
\end{equation}
This is the general form of the phase of any self-similar solution
to Eq. (\ref{NLS}) with amplitude profile given by $\rho$.
However, not all of these solutions correspond to physically
interesting ones. To understand why let us compute
\begin{equation}
K(t) \equiv \int_{\mathbb{R}^n} |\nabla u|^2 d^nx =
\int_{\mathbb{R}^n} \left[\rho_r^2 + \rho^2 \phi_r^2 \right] d^nx.
\end{equation}
As discussed above this integral (which is related to the $H^1$
norm of $u$) must be finite. However, it is clear that
\begin{eqnarray}
K(t) & \geq & \int_{\mathbb{R}^n} \rho^2 \phi_q^2 d^nq =
\int_0^{\infty} \rho^2 \left[\frac{C_2(t)}{q^{n-1}\rho(q)^2} +
\frac{q}{2}\right]^2q^{n-1}dq \int_{\mathbb{S}^n} d\Omega
\nonumber \\& =& S_n \left[ \int _0^{\infty} C_2(t)
\left(\frac{C_2(t)}{q^{n-1}\rho(q)^2} + q\right) dq + \int
_0^{\infty} q^{n+1}\rho(q)^2 dq\right] \label{17}
\end{eqnarray}
where $\mathbb{S}^n$ denotes the surface of the $n$-dimensional
sphere of unit radius and $S_n = \int_{\mathbb{S}^n} d\Omega$ is
its measure. The most singular term in Eq. \ref{17} is the first
one. Taking into account that $\int_{\mathbb{R}^n} \rho^2 d^nq =
1$ we can use the Schwartz inequality to prove the divergence of
the first term of Eq. (\ref{17})
\begin{multline}
\left( \int_{\mathbb{R}^n} \rho^2 d^nq \right)
\left(\int_0^{\infty} \frac{C_2(t)}{q^{n-1}\rho(q)^2} dq
\int_{\mathbb{S}^n} d\Omega\right) \\ = \left( \int_0^{\infty}
\rho^2 q^{n-1} dq \right) \left(\int_0^{\infty}
\frac{C_2(t)}{q^{n-1}\rho(q)^2} dq\right) S_n^2 \\ \geq S_n^2
\int_{0}^{\infty} \rho^2 q^{n-1} \frac{C_2(t)}{q^{n-1}\rho(q)^2}
dq = C_2(t) S_n^2 \int_0^{\infty} dq \rightarrow \infty.
\end{multline}
Which proves the divergence of $K(t)$ unless $C_2(t) = 0$. The
second term in \eqref{17} cannot compensate this divergence since
this would require $\rho^2 \sim 1/q^n$ for $q \rightarrow \infty$
which would lead to non-normalized solutions. Using this fact and
changing back to physical variable $r$ we get finally for the
phase
\begin{equation}\label{phasesf}
\phi(r,t) = \frac{\dot{w}}{2w}r^2.
\end{equation}
Eq. (\ref{phasesf}) is one of the main results of this paper:
\emph{the phase of finite-energy self-similar solutions is
cuadratic} irrespective of the many possible variations of the
problem (nonlinearity, potential, spatial dimensionality). This
fact explains why the choice of quadratic phases as in \eqref{ccc}
in time-dependent variational method leads to very good results.
This fact also clarifies why these approximate methods lead
usually to a prefactor $\beta = \dot{w}/(2w)$, which is indeed the
exact result.

We must emphasize that phase choices such as \eqref{phasesf} have
been used previously by other authors either as approximate
expressions for the phase as comented before or as an specific
choice leading to particular solutions (e.g. in the framework of
the theory of self-focusing in NLS equations \cite{Sulem,Fibich}).
What we prove here is that \emph{all finite-energy self-similar
solutions to Eq. (\ref{NLS}) must have a cuadratic phase}, which
is a stronger result.

\subsection{Equations for the scaling parameter}

When self-similar solutions (or quasi-solutions in the sense
discussed above) are considered it is possible to get closed
equations for the evolution of the most relevant parameter $w(t)$
which determines the width of the solutions. In this section we
restrict ourselves to the autonomous case when $V$ and $g$ do not
depend explicitly on $t$. Let us first consider the hamiltonian
\begin{equation}
\label{Ham} H = \frac{1}{2} \int |\nabla u|^2 d^nx + \int G(a^2)
d^nx + \int V(x) |u|^2 d^nx,\end{equation} where $G$ is a real
function satisfying $g = \partial G/\partial (a^2)$. It is easy to
check that $H$ is a conserved quantity which is assumed to be
finite. For self-similar solutions of the form (\ref{selfsimm})
and using again the definition $q = r/w$, we get
\begin{eqnarray}
\nonumber H & = & \frac{1}{2} \int_{\mathbb{R}^n}
\left(\left|\frac{\partial \rho}{\partial r}\right|^2 + \rho^2
\left|\frac{\partial \phi}{\partial r}\right|^2\right) d^n x
+\int_{\mathbb{R}^n} G(A^2\rho^2) d^nx
+ A^2 \int_{\mathbb{R}^n} V(x) \rho^2 d^nx \\
& = & \frac{1}{2} A^2 w^n \int_{\mathbb{R}^n}
\left(\frac{\rho_r^2}{w^2} + \frac{\rho^2}{w^2} \phi_q^2\right)
d^nq + w^n \int_{\mathbb{R}^n} G\left(\frac{\rho^2}{w^n}\right)
d^nq \nonumber \\
& & + \int_{\mathbb{R}^n} A^2 w^n V\left(wq\right) \rho^2 d^nq.
\end{eqnarray}
Using the $L^2$-norm constraint \eqref{normac} we get
\begin{eqnarray} \nonumber
H & = & \frac{1}{2w^2} \int_{\mathbb{R}^n} \rho_q^2 d^n q +
\frac{1}{2w^2} \int_{\mathbb{R}^n} \rho^2 \phi_q^2 d^nq  \\
& & + w^n \int_{\mathbb{R}^n} G\left(\rho^2/w^n\right) d^nq +
\int_{\mathbb{R}^n} V\left(wq\right) \rho^2 d^nq.
\end{eqnarray}
Finally, we use \eqref{phasefin} and the fact that
\begin{equation}
\int_{\mathbb{R}^n}d^nq = \int_0^{\infty}
q^{n-1}dq\int_{\mathbb{S}^n}d\Omega \equiv S_n\int_0^{\infty}
q^{n-1}dq, \end{equation}
 to get
\begin{eqnarray} \nonumber \frac{H}{S_n} & = & \frac{1}{2w^2} \int_0^{\infty} \rho_q^{2}q^{n-1} dq +
\frac{1}{2}\dot{w}^2 \int_{0}^{\infty} \rho^2 q^{n+1} dq  \\
& & + w^n \int_0^{\infty} G\left(\rho(q)^2/w^n\right) q^{n-1} dq +
\int_0^{\infty} V\left(wq\right) \rho^2 q^{n-1} dq.\label{Hfin}
\end{eqnarray}
Differentiating \eqref{Hfin} with respect to time we obtain
\begin{multline}
-\frac{\dot{w}}{w^3}\int_0^{\infty} \rho_q^2 q^{n-1}dq +
\dot{w}\ddot{w} \int_0^{\infty} q^{n+1} \rho^2 dq +n w^{n-1}
\dot{w} \int_0^{\infty} G\left(\rho^2/w^n\right) q^{n-1}dq
\\ - w^n \int_0^{\infty} \frac{nw^{n-1}\dot{w}}{w^{2n}} \rho^2 g\left(\rho^2/w^n\right)q^{n-1}dq +
\int_0^{\infty} V'(wq)q\dot{w}\rho^2q^{n-1}dq=0,\label{multiw}
\end{multline}
which can be written as a Newton-like equation
\begin{subequations}\label{Newton}
\begin{eqnarray}
\left(\int_0^{\infty}q^{n+1}\rho^2\right) \ddot{w} & = &
\frac{\left(\int_0^\infty q^{n-1}\rho_q^2 dq\right)}{w^3} \label{Newtona} \\
& & - nw^{n-1} \int_0^{\infty} \left[
G\left(\frac{\rho^2}{w^n}\right)- \frac{\rho^2}{w^n} g
\left(\frac{\rho^2}{w^n}\right)\right] q^{n-1}dq \label{Newtonb}\\
& & - \int_0^{\infty} q^n V'(wq)\rho^2 dq.\label{Newtonc}
\end{eqnarray}
\end{subequations}
The different terms in Eq. (\ref{Newton}) have a very clear
physical interpretation. The right-hand-side of Eq.
(\ref{Newtona}) is a repulsive term proportional to $1/w^3$ which
accounts for the tendency of the wave-packet to spread under the
action of dispersion (this is mathematically described by the term
with the Laplacian in Eq. (\ref{NLS}). The second one (Eq.
\eqref{Newtonb}) is the self-interaction energy due to the
nonlinearity. Finally the last term given by Eq. (\ref{Newtonc})
comes from the interaction with the external potential $V$.

Eqs. (\ref{Newton}) are of the same form as the equations arising
in particular studies of Eq. (\ref{NLS}) based on the method of
collective coordinates of which Eqs. (\ref{Newton}) are a
generalization. From them it is possible to understand that when
self-similar solutions exist what the variational method does is
to fit somehow the right dynamics but using different values for
the integrals depending on $\rho$. Due to the lack of knowledge of
the right self-similar profiles $\rho$, the integrals are
estimated by using the ansatz $\varphi$. If the ansatz is
constructed reasonably, e.g. taking into account the right
asymptotic behavior of the solutions, we may expect only a
quantitative discrepancy between the exact results for
self-similar solutions and the estimates obtained from the
collective coordinate method obtained from Eq. (\ref{Newton}).

  The  main advantage of our derivation of Eqs. (\ref{Newton})
 is that they are obtained systematically on the basis of
the exact phase of self-similar solutions. These equations provide
then a direct way to obtain estimates for the dynamics when
self-similar solutions do not exist and allow to avoid the lengthy
calculations of the method of collective coordinates.

There is a lot of information contained in Eqs. (\ref{Newton}).
Specifically, they describe many cases previously studied. For
instance, the simplest case $g(y) = g_0y^2, g_0 <0, V = 0$
corresponds to the usual focusing cubic NLS. From there we obtain
immediately the following condition for the collapse of
self-similar solutions
\begin{equation}
g_0 < g_c = -\frac{\int_0^{\infty} (\rho_q)^2 q^{n-1}
dq}{\int_0^{\infty} q^{n-1} \rho^4 dq}
\end{equation}
which allows to estimate the critical value of $g$ for solutions
with $\|\rho\|_2=1$. For instance taking gaussians one gets $g_c =
-2\pi$, which is the usual estimate obtained from time-dependent
variational methods. Taking $\rho$ to be the so-called
Townes-soliton \cite{Sulem} leads to the exact result for the
critical value of $g$ \cite{Sulem}.

Many other situations have been described in the literature.
Taking $V(r) = \tfrac{1}{2} \Omega^2 r^2$ and $g(y) = g_0y^2, g_0
\in \mathbb{R}$ we get the simplest situation of Bose-Einstein
condensation in a symmetric trap. This problem was studied
variationally in Ref. \cite{Perez97}. Our equations coincide with
those for the spatially symmetric case discussed there. Also in
Refs. \cite{cc1,cc2,cc3} cubic-quintic nonlinearities where used
to describe four-body collisions in Bose-Einstein condensates.
Eqs. (\ref{Newton}) give us the evolution equations found there by
just choosing a nonlinearity of the type $g(y) = g_3y^2 + g_5y^4$.

\section{Conclusions and discussion}
\label{VI}

In this paper we have studied the phase of self-similar solutions
to general Nonlinear Schr\"odinger equations. Our analysis
provides a better understanding of the way collective coordinate
methods work. Also we have used the results on the phase to find
an effective equation describing exactly the dynamics of
self-similar solutions to the NLS equation and approximating the
dynamics of other, more complicated solutions.

Our analysis on the phase can be easily extended to more
complicated vector systems. We expect that the procedure followed
here may provide a basis for choosing systematically the phase for
more complex ansatzs such as those which are not self-similar thus
overcoming one of the difficult points of usual collective
coordinate methods. It is worth mentioning that our analysis of
the phase is applicable to non-autonomous situations such as the
ones recently discussed in Refs. \cite{new1,new2,new3,new4,new5},
which represent an emerging field of applications of Nonlinear
Sch\"odinger Equations.

 \textbf{Acknowledgments}

This work has been partially supported by the Ministerio de
Ciencia y Tecnolog\'{\i}a and FEDER under grants BFM2000-0521 and
BFM2003-02832 and Consejer\'{\i}a de Ciencia y Tecnolog\'{\i}a de
la Junta de Comunidades de Castilla-La Mancha under grant
PAC02-002.


\end{document}